\documentclass[twocolumn,article,aps,prb,showpacs]{revtex4}
\usepackage{bm,epsfig,psfig,graphicx,color}
\begin{document}

\title{{\bf Optical Spectra and Localization of Excitons in Inhomogeneous \\
Helical Cylindrical Aggregates}}
\author{C{\u{a}}t{\u{a}}lin Didraga
and Jasper Knoester\footnote[1]
        {Corresponding author. Fax: 31-50-3634947. E-mail:
        knoester@phys.rug.nl}}

\affiliation
{Institute for Theoretical Physics and Materials Science
Centre, University of Groningen, Nijenborgh 4, 9747 AG  Groningen,
The Netherlands}

\date{\today}

\begin{abstract}

We study the linear optical properties of helical cylindrical
molecular aggregates accounting for the effects of static diagonal
disorder. Absorption, linear dichroism, and circular dichroism
spectra are presented, calculated using brute force numerical
simulations and a modified version of the coherent potential
approximation that accounts for finite size effects. Excellent
agreement between both approaches is found. It is also shown that
the inclusion of disorder results in a better agreement between
calculated and measured spectra for the chlorosomes of green
bacteria as compared to our previous report, where we restricted
ourselves to homogeneous cylinders [J. Phys. Chem. B {\bf 106},
11474 (2002)]. We also investigate the localization properties of
the excitons responsible for the optical response. By analyzing an
autocorrelation function of the exciton wave function, we find a
strongly anisotropic localization behavior, closely following the
properties of chiral wave functions which previously have been
found for homogenoeus helical cylinders [J. Chem. Phys. {\bf 121},
946 (2004)]. It is shown that the circular dichroism spectrum may
still show a strong dependence on the cylinder length, even when
the exciton wave function is localized in a region small compared
to the cylinder's size.

\end{abstract}

\pacs{ 71.35.Aa;
                        78.30.Ly;
71.35.Cc  
                        78.67.-n
}

\maketitle

\section{Introduction}\label{Introduction}

The optical properties and optical dynamics of molecular
aggregates with a cylindrical geometry currently draw considerable
attention. Both natural and synthetic forms of such molecular
nanotubes are investigated. Among the natural systems, the rod
shaped light-harvesting complexes in the chlorosomes of green
bacteria are well-known examples.\cite{STAEHELIN} The chlorosomes
of {\it{Chloroflexus aurantiacus}} contain tens of thousands of
bacteriochlorophyll molecules self-assembled in cylindrical
structures with a monolayer wall of roughly $5\,$nm diameter and a
length of hundreds of
nanometers.\cite{HOLZWARTH,BALABAN,ROSS,MIZOGUCHI} The
light-harvesting system of the bacterium {\it{Chlorobium tepidum}}
also contains cylindrical aggregates, with a bilayer wall and a
diameter of roughly $10\,$nm.\cite{VANROSSUM} These natural
systems should be referred to as J aggregates, as the absorption
spectrum is red-shifted relative to the transition frequency of a
single bacteriochlorophyll molecule.

J aggregates with a cylindrical geometry have recently also been
prepared via synthetic routes. In particular a class of
substituted 5,5',6,6'-tetrachlorobenzimidacarbocyanine dyes has
been created that forms such aggregates; the cylindrical geometry
was revealed using cryo-TEM.\cite{KIRST,BERL1} It has been
demonstrated that the precise morphology as well as the details of
the optical properties depends on the nature of the substituents
and the solvent.\cite{BERL3,C8S3} These synthetic cylinders
usually have bilayer walls with an outer diameter of about
$15\,$nm and a wall thickness of $4\,$nm. The cylinder length
extends to several hundreds of nanometers. Recently, the (helical)
arrangement of the molecules inside the bilayer wall of such
aggregates has been determined for the first time. This was done
for the aggregates of the dye
3,3'-bis(3-sulfopropyl)-5,5',6,6'-tetrachloro-1,1'-dioctylbenzimidacarbocyanine
(C8S3), by modeling the data from cryo-TEM, absorption and linear
dichroism measurements.\cite{C8S3} A Frenkel exciton model based
on two weakly interacting bricklayer monolayers wrapped on
cylindrical surfaces of appropriate diameter gave a good fit to
experiment. Very recently, it has been discovered that bilayer
molecular nanotubes (diameter 14 nm) may also be formed through
self-assembly of amphiphilic hexi-{\em peri-}hexabenzocoronene
molecules.\cite{HILL04} Another class of synthetic
cylindrical aggregates are those formed through self-assembly of
the dye meso-tetra(4-sulfonatophenyl)porphyrin (TTPS$_4$) in acidic
aqueous solution.\cite{GANDINI,ROTOMSKIS} From small angle x-ray
scattering (SAXS)\cite{GANDINI} and atomic force microscopy
(AFM)\cite{ROTOMSKIS} it has been concluded that these aggregates
are hollow monolayer tubes with a diameter of about 20 nm.
Finally, it has been shown that under the influence of tetrahedral
chemical defects conjugated polymers may also adopt ordered
cylindrical conformations.\cite{BARBARA}

In previous model studies of cylindrical aggregates, we have
mainly restricted ourselves to homogeneous aggregates, i.e., we
have ignored the role of disorder on the Frenkel exciton states
that determine the optical response.\cite{BEDNARZ01,SPITZ2,DIDR,DIDR04}
 An important simplification
occurs in this case, as the cylindrical symmetry may then be used
to distinguish excitons in classes (bands) of different transverse
quantum number $k_2$, which describes the Bloch nature of the
exciton wave function in the direction along the circumference of
the cylinder.\cite{DIDR} The exciton eigenstates may then be
determined from a set of one-dimensional effective Hamiltonians
for each value of $k_2$. Importantly, the introduction of the
transverse quantum number dictates simple selection rules, which
state that only states with $k_2=0$ or $k_2= \pm 1$ can be
observed in linear optics. States in the $k_2=0$ band give rise to
absorption polarized along the cylinder's axis, while states with
$k_2=\pm 1$ yield a polarization perpendicular to this axis. For
each of these three bands only a few strongly allowed
(superradiant) exciton states occur.\cite{DIDR,DIDR04}

In spite of the fact that usually appreciable energetic or
interaction disorder occur in self-assembled molecular aggregates,
it turns out that a homogeneous aggregate model, with the simple
selection rules discussed above, does describe the salient
features of experimentally observed spectra. For instance, using
this model, the experimentally observed variation in the CD
spectra of the chlorosomes of {\it{Chloroflexus aurantiacus}} was
explained.\cite{DIDR} Similarly, the polarization dependent
spectra of the bilayer C8S3 cylinders are rather well described
neglecting the role of disorder.\cite{C8S3} We have shown,
however, that the inclusion of static energy disorder does improve
the comparison to experiment for the latter case.\cite{C8S3}
Similarly, model results for chlorosomes that do account for
disorder also show a better quantitative comparison to
experiment.\cite{PROKHO_NEW}

The aim of this paper is to systematically study the effect of
static diagonal disorder on the optical spectra of cylindrical
molecular aggregates and to study the exciton localization
properties caused by the presence of disorder. While the formalism
used applies to general cylindrical aggregates, we will in
explicit calculations restrict ourselves to cylinders of the
structure of chlorosomes of {\it{Chloroflexus aurantiacus}}. As
observable quantities, we will focus on the absorption, linear
dichroism (LD), and circular dichroism (CD) spectra. The disorder
averages of these specta are calculated using numerical
simulations as well as a modified version of the coherent
potential approximation (CPA) that accounts for finite-size
effects.

Localization properties have been extensively studied for Frenkel
excitons in linear \cite{SCHREIBER1,SCHREIBER2,FIDDER,MALY01} and
circular \cite{CHACH,MEIER,KUHN} aggregates. For the higher
dimensional cylindrical aggregates, such studies have not been
performed yet. We will fill this gap by investigating the inverse
participation ratio and an autocorrelation function of the wave
function. Using the latter, we will show a strong anisotropy in
the localization properties of the excitons, which may be traced
back to the recently discovered chiral behavior of the exciton
wave functions on finite homogeneous cylindrical
aggregates,\cite{DIDR04} dictated by the behavior of the dipolar
excitation exchange interactions in the system. We will also
demonstrate that the CD spectrum may exhibit a dependence on the
cylinder's length, even when the excitons are localized on regions
small compared to the its size.

The outline of this paper is as follows. In Sec.~\ref{model} we
will present the exciton model and give the general expressions
for the quantities of interest, in particular for the spectra and
the localization characteristics. Section \ref{CPA} is dedicated to
explain the modified version of the CPA. Results for the spectra
are presented and discussed in Sec.~\ref{results}, while in
Sec.~\ref{results2} we do the same for the localization
characteristics. Finally, we present our conclusions in
Sec.~\ref{conclusions}.

\section{Model and quantities of interest}\label{model}

\paragraph{Aggregate structure, Hamiltonian, and eigenstates.}\label{eigenstates}

For the structure of the cylindrical aggregate we will use the
same model and notation as described in Ref.~\onlinecite{DIDR} and
depicted in Figure~1. The aggregate consists of a stack of $N_1$
rings of radius $R$ (labeled $n_1=1,\ldots, N_1$), each containing
$N_2$ molecules. The distance between neighboring rings is denoted
$h$. Neighboring rings are rotated relative to each other over a
helical angle $\gamma$, with $0 \le \gamma < 2\pi/N_2$. Connecting
the closest molecules on neighboring rings, one observes that the
aggregate may be viewed as $N_2$ helices each containing $N_1$
molecules, winding around the cylinder (the dashed line in Figure~1
indicates one such helix). Each molecule may now be labeled ${\bf
n}=(n_1,n_2)$, where $n_1$ denotes the ring on which the molecule
resides, while $n_2$ denotes the helix on which it lies. The total
number of molecules in the aggregate is denoted $N=N_1 N_2$. In
Ref.~\onlinecite{DIDR} we explained the general nature of this
structural model.

\begin{figure}[tb]
         \centerline{
           \scalebox{1.0}{
             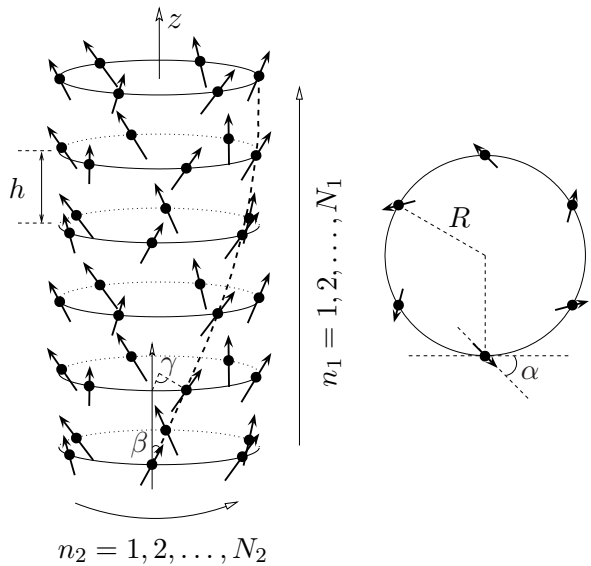
             }
           }
         \caption{Cylindrical aggregate consisting of a stack of
$N_1$ rings, labeled $n_1=1,2,\ldots,N_1$, that each contain $N_2$
molecules, labeled $n_2=1,2,\ldots,N_2$. The arrows indicate the
transition dipoles, which are equal in magnitude ($\mu$) and make
an angle $\beta$ with the cylinder axis.  The projection of each
dipole on the plane of the rings makes an angle $\alpha$ with the
local tangent to the ring (see projection of one ring displayed to
the right). Each ring is rotated with respect to the previous one
over an angle $\gamma$, so that we may view the aggregate as a
collection of $N_2$ parallel helices on the cylinder's surface.
One such helix is indicated by the dashed curve. The label $n_2$
in fact labels the helices.}
       \end{figure}

All molecules are modeled as identical two-level systems, with
transition dipoles $\bm{\mu_{\mathrm n}}$ that are equal in
magnitude ($\mu$) and have equal orientations relative to the frame of
the cylinder at the position of the molecule.  In particular, all
molecular dipoles make an angle $\beta$ with the cylinder axis
(referred to as the $z$ axis), while the projection of each dipole
on the $xy$-plane makes an angle $\alpha$ with the local tangent
to the rings. Explicitly, the $x$, $y$, and $z$ components of the
molecular position vectors and dipole moments are given by the
three-dimensional vectors:
\begin{equation}
{\mathbf {r_n}}=\left( R \cos (n_2\phi_2+n_1\gamma), R
\sin(n_2\phi_2+n_1\gamma),n_1 h\right)
\end{equation}
and
\begin{eqnarray}
\nonumber \bm{\mu_{\mathrm n}} &=& \left( -\mu \sin\beta \sin(n_2\phi_2+n_1\gamma
 -\alpha), \right. \\
  & & \left. \mu \sin\beta
 \cos(n_2\phi_2+n_1\gamma-\alpha),\mu\cos\beta\right)\,,
\end{eqnarray}
respectively, with $\phi_2=2\pi/N_2$.

The electronically excited states of the aggregate are described by
the Frenkel exciton Hamiltonian with static diagonal disorder.
Setting $\hbar=1$, we have
\begin{equation}\label{Hamiltonian}
H=\sum_{\mathbf n}(\omega_0+\epsilon_{\mathbf n}) b_{\mathbf n}^\dagger b_{\mathbf n} +
{\sum_\mathbf{n,m}}' J(\mathbf{n-m}) b_{\mathbf n}^\dagger
b_{\mathbf m}\,,
\end{equation}
where $b_{\mathbf n}^{\dagger}$ and $b_{\mathbf n}$ denote the
Pauli operators for creation and annihilation of an excitation on
molecule ${\mathbf n}$,
respectively.\cite{Agranovich59,Agranovich61,DAVY,Agranovich}
Furthermore, $\omega_0$ is the average molecular transition
frequency and $\epsilon_{\mathbf n}$ is the static random energy
offset at site ${\bf n}$, induced by slow solvent effects.
$J(\mathbf{n-m})$ is the excitation transfer interaction between
molecules ${\mathbf n}$ and ${\mathbf m}$.  Due to the symmetry of
the system the interaction only depends on the relative positions
of the two molecules.  The prime on the summation indicates that
the term with ${\mathbf n}= {\mathbf m}$ is excluded from the
summation. We assume that $J(\mathbf{n-m})$ results from
dipole-dipole interactions, giving it the explicit form
\begin{equation}\label{V}
J(\mathbf{n-m})=\frac{\bm{\mu_{\mathrm n}}\cdot \bm{\mu_{\mathrm
m}}} {\left|\mathbf{r_{nm}}\right|^3} -3\frac{(\bm{\mu_{\mathrm
n}}\cdot \mathbf{r_{nm}}) ( \bm{\mu_{\mathrm m}}\cdot
\mathbf{r_{nm}})} {\left|\mathbf{r_{nm}}\right|^5} \,, \label{J}
\end{equation}
with  $\mathbf{r_{nm}}=\mathbf{r_n}-\mathbf{r_m}$.

The Hamiltonian Eq.~(\ref{Hamiltonian}) differs from the one
discussed in Ref.~\onlinecite{DIDR} only in the inclusion of random
energy offsets $\epsilon_{\mathbf n}$. We will assume that the
energy offsets on different molecules are uncorrelated and follow
a Gaussian distribution ${\mathcal P}(\epsilon_{\mathbf n})$ with
standard deviation $\sigma$. Hence, each $\epsilon_{\mathbf n}$ is
taken independently from the distribution
\begin{equation}\label{distri}
{\mathcal P}(\epsilon_{\mathbf
n})=\frac{1}{\sqrt{2\pi}\sigma}\exp\left(-\frac{\epsilon_{\mathbf
n}^2}{2\sigma^2} \right).
\end{equation}

To describe the linear optical response of the aggregate, it
suffices to consider the space of one-exciton states, i.e., those
states in which the molecules of the cylinder share one
excitation. The general form of these eigenstates reads
\begin{equation}\label{Eigenstate}
|q\rangle = \sum_{\bf n} \varphi_q({\bf n}) b_{\bf n}^{\dagger}
|g\rangle, \label{state}\,
\end{equation}
where $|g\rangle$ denotes the overall ground state, in which all
molecules are in their ground state. We have used the label $q$ to
distinguish the $N$ one-exciton states; the quantity
$\varphi_q({\bf n})$ denotes the amplitude of the $q$th state on
molecule ${\bf n}$. These amplitudes are obtained by diagonalizing
the $N \times N$ one-exciton Hamiltonian, which has the quantities
$\omega_0+\epsilon_{\mathbf n}$ as diagonal elements and the
$J(\mathbf{n-m})$ as off-diagonal ones. The eigenvectors
$\varphi_q({\bf n})$ will be assumed to be normalized to unity.
Unless stated otherwise, we will impose open boundary conditions
along the $z$ axis.

In the absence of disorder ($\sigma=0$), the label $q$ may be
replaced by a two-dimensional label $\mathbf{k}=(k_1,k_2)$, where
$k_2$ denotes the wave number describing the Bloch momentum of the
exciton state along the ring direction, while $k_1$ labels the
$N_1$ possible exciton states in each of the $N_2$ different bands
characterized by one value of $k_2$.\cite{DIDR,DIDR04} The
diagonalization then separates into $N_2$ independent effective
one-dimensional problems. In the current general case of disorder,
we cannot make this decomposition of the quantum labels, and we
will keep the general label $q$. We note that with the breakdown
of the separation into transverse ($k_2$) and longitudinal ($k_1$)
quantum numbers, also the strict distinction between states
polarized parallel ($k_2=0$) and perpendicular ($k_2=\pm 1$) to the
$z$ axis breaks down. In general, the dipoles of the exciton
states in disordered cylinders may have any orientation relative
to the $z$ axis.

\paragraph{Optical spectra.}\label{opt_spectra}
We will be interested in calculating the absorption, LD, and CD
spectra in the presence of diagonal disorder. The general
expressions for these spectra in terms of the one-exciton energies
$E_q$ and eigenvector components $\varphi_q(\mathbf{n})$ follow
from linear response theory and take the generic form\cite{DIDR}
\begin{equation}\label{SPECTRA}
S(\omega)=\left\langle \sum_{q} X_{q} \delta(\omega-E_{q})
\right\rangle\,,
\end{equation}
with strength
\begin{equation}\label{Xq}
X_q=\sum_{\mathbf{n,m}}  \varphi_q(\mathbf{n})
\varphi^*_q(\mathbf{m}) X_{\mathbf{n,m}}\,.
\end{equation}
Here, $S(\omega)$ stands for $A(\omega)$, $LD(\omega)$, and
$CD(\omega)$ in case of the absorption, LD, and CD spectrum,
respectively, and the angular brackets $\langle\ldots\rangle$
denote the average over the random energy offsets. Furthermore, the
quantities $X_{\mathbf{n,m}}$ are the corresponding strengths in
the site representation, which take the form
\begin{equation}
\label{XA} X_{\mathbf{n,m}}^{A}= \frac{1}{3}\mu^2\cos^2{\beta} +
\frac{1}{3}\mu^2\cos{\xi} \sin^2{\beta}\,,
\end{equation}
\begin{equation}
\label{XLD} X_{\mathbf{n,m}}^{LD} = \mu^2\cos^2{\beta} -
\frac{1}{2}\mu^2\cos{\xi} \sin^2{\beta}\,,
\end{equation}
and
\begin{eqnarray}
\nonumber \label{XCD} X_{\mathbf{n,m}}^{CD} &=&
\frac{\pi\mu^2}{6\lambda}\left[R(1-\cos{\xi})\sin{(2\beta)}\cos{\alpha} \right. \\
 & & \left. -(n_1-m_1)h\sin{\xi}\sin^2{\beta}\right].
\end{eqnarray}
for the three spectra considered. Here,
$\xi\equiv[(n_2-m_2)\phi_2+(n_1-m_1)\gamma]$ and $\lambda$ denotes
the wavelength of the light. In the above expressions for the
absorption and CD spectra, an isotropic average over orientations
of the cylinder has been used (appropriate for an isotropic
solution), while in case of the LD spectrum, a uniform average over
angles of rotation around the cylinder's axis was inferred
(appropriate for samples with perfect alignment of the cylinders).
Any (unlikely) correlation between orientation and disorder
realization has been neglected.  For future reference it is useful
to note that also the density of states, $\rho({\omega})$, follows
the generic expression Eq.~(\ref{SPECTRA}), with
$X_{\mathbf{n,m}}^{\rho}=\delta_{\mathbf{n,m}}$, implying
$X_q^{\rho}=1$.

The disorder averaged spectra and density of states in principle
may be calculated using straightforward numerical simulations, in
which one generates a number of random disorder realizations
$\{\epsilon_{\mathbf n}\}$ and for each realization performs an
explicit diagonalization of the one-exciton Hamiltonian to obtain
the quantities $E_q$ and $\varphi_q(\mathbf{n})$. For
one-dimensional (linear or circular) molecular aggregates this is
common practice. For higher-dimensional aggregates, however, such
as planes or cylinders, this method may be quite demanding, due to
the large number of molecules involved in these systems and the
tendency for the collective excitations to be more delocalized in
higher dimensions. This motivates the use of alternative, albeit
approximate, methods to calculate the average spectra. A
well-known example is the coherent potential approximation (CPA),
\cite{SOVEN,TAYLOR} which previously has been applied with success
to calculate spectra of, for instance, isotopically mixed aromatic
crystals (dichotomic disorder),\cite{HOSHEN,PORT} one-component
systems with Gaussian diagonal disorder,\cite{HUBER1,HUBER2} and
two-component systems with bi-Gaussian diagonal
disorder.\cite{BAK02}

In this paper, we will use both brute-force numerical simulations
and the CPA to calculate the spectra and the density of states. We
will extend the usual CPA to account for finite-size effects
(Section~\ref{CPA}) and show that this modified CPA gives excellent
agreement with numerical simulations, in fact significantly better
than its conventional implementation.

\paragraph{Localization characteristics.}
As is well known, the presence of diagonal disorder leads to the
localization of the excitonic eigenstates.\cite{ANDERSON} In order
to have quantitative information on this localization, we will
analyze the inverse participation ratio and an autocorrelation
function of the wave function. Both quantities can only be
addressed within a numerical simulation. The energy dependent
inverse participation ratio is defined
through\cite{THOULESS,SCHREIBER1,SCHREIBER2,FIDDER}
\begin{equation}\label{PARRATIO}
{\mathcal L}(\omega)= \left\langle \sum_{q} \left[ \sum_{\mathbf
n}|\varphi_{q}({\mathbf n})|^4 \delta(\omega-E_{q}) \right]
\right\rangle /\rho(\omega)\,.
\end{equation}
The participation ratio, ${\mathcal L}^{-1}$, is generally
accepted as a typical value for the number of molecules
participating in the eigenstates at energy $\omega$. For example,
a state localized on a single molecule has ${\mathcal
L}(\omega)=1$, whereas for the completely delocalized states on a
homogeneous cylinder ${\mathcal L}(\omega) \sim 1/N$. In the
latter case, the precise value depends on the boundary conditions
and on the possible combination of degenerate complex eigenstates
to real ones. If we use periodic boundary conditions along the $z$
axis and use real $\sin$ and $\cos$ forms for the transverse and
longitudinal Bloch wave functions, all states (except a
few\cite{footnote}) have a participation ratio $9/(4N)$. For the
same boundary conditions using complex wave functions, all states
have ${\mathcal L}(\omega) = 1/N$. If we use open boundary
conditions with real transverse wave functions and accept the
ansatz solutions for the longitudinal wave functions analyzed in
Ref.~\onlinecite{DIDR04}, we obtain (again with a few exceptions)
the value $9/[4(N_1+1)N_2] \approx 9/(4N)$.

Alternative measures for the extent of the exciton wave function
have also been considered, for instance, autocorrelation functions
of the wave function were used to study the ring-shaped LH2 antenna
system.\cite{CHACH,KUHN_COH} Correlation functions are
particularly useful when dealing with anisotropic
higher-dimensional systems, such as cylindrical aggregates, as
they allow for a study of the localization properties along
different spatial directions.  Such information cannot be
extracted from the inverse participation ratio. Hence, we define
the autocorrelation function
\begin{equation}
\label{COH} {\mathcal C}({\mathbf n};\omega)=\left\langle \sum_{q}
\sum_{\mathbf m}\left|\varphi_{q}({\mathbf m})
\varphi_{q}^*(\mathbf{m+n})\right|\delta(\omega-E_{q})
\right\rangle/\rho(\omega),
\end{equation}
where the summation over ${\mathbf m}$ extends over $m_1=1,\ldots
, N_1-n_1$ and $m_2=1,\ldots , N_2$ in order to be consistent with
open boundary conditions in the $n_1$ direction.

The generic form of ${\mathcal C}({\mathbf n};\omega)$ on the
$(n_1,n_2)$ plane is a structure that peaks at the origin $(0,0)$
(where it has the value unity) and which (after averaging over a
sufficient number of disorder realizations) has inversion symmetry
with respect to the origin. The form of the peak shows in what
direction the exciton wave functions at energy $\omega$ are most
localized or extended. Finally, from the autocorrelation one can
define a localization measure (alternative to ${\mathcal
L}^{-1}(\omega)$) for the total number of molecules participating
in the typical wave function at energy $\omega$. We will denote
this measure by $N_{del}^C(\omega)$ and define it as the total
number of ${\bf n}$ values (molecules) with ${\mathcal C}({\mathbf
n};\omega) > {\mathcal C}({\bf 0};\omega)/e=1/e$ ($e$ the base of
the natural logarithm). It should be stressed that both the
participation ratio and $N_{del}^C(\omega)$ represent typical
numbers and cannot be expected to give exactly the same result.
However, one does expect these two measures to vary in a similar
way with energy, disorder strength, or system size. We will come back
to this in Sec.~\ref{results2}.

\section{Modified Coherent Potential Approximation} \label{CPA}

In this section, we address some essential technical aspects to
use the CPA when calculating the optical spectra. The method as such
is well-documented in text books,\cite{RICKAYZEN,ECONOMOU} which
is why we only focus on two aspects that are specific to our
application. The first one concerns reducing the general
expression Eq.~(\ref{SPECTRA}) for the spectra to a form that can
be addressed within the CPA. The second one concerns the treatment
of finite systems, where periodic boundary conditions should be
avoided.

The CPA is a method that yields an approximate form for the
disorder averaged (retarded) Green's function $\left\langle \hat
G(\omega)\right\rangle$, with
\begin{equation}
\hat G(\omega)=(\omega\hat 1- H +{i} \eta)^{-1}\,.
\end{equation}
Here, $\hat 1$ is the unit operator, $\eta$ is a positive
infinitesimal constant, and $H$ is the Hamiltonian
Eq.~(\ref{Hamiltonian}). Using the Green's function, we may rewrite
Eq.~(\ref{SPECTRA}) for the spectra as
\begin{equation}
\label{SPECTRAG}
S(\omega)=-\frac{1}{\pi}\mathrm{Im}\left\langle\sum_q X_q \langle
q| \hat G(\omega)|q\rangle\right\rangle\,.
\end{equation}
As the eigenstates $|q\rangle$ and the strengths $X_q$ depend on
the disorder realization, some care is needed to reduce
Eq.~(\ref{SPECTRAG}) to a form that only contains $\left\langle
\hat G(\omega)\right\rangle$.

We first use Eq.~(\ref{Xq}) for $X_q$ in terms of
$X_{\mathbf{n,m}}$. Realizing that $\varphi_q(\mathbf{n})=\langle
\mathbf{n}|q\rangle$ and using the fact that the $X_{\mathbf{n,m}}$
do not depend on the disorder realization [see Eqs.~(\ref{XA}),
(\ref{XLD}), and (\ref{XCD})], we may rewrite
\begin{equation}\label{Snm}
S(\omega)=-\frac{1}{\pi}\mathrm{Im}\sum_{\mathbf{n,m}}
X_{\mathbf{n,m}} \langle \mathbf{n}|
\left\langle\hat G(\omega)\right\rangle|\mathbf{m}\rangle.
\end{equation}
We now change to the basis of eigenstates of the system in the
absence of disorder ($\sigma=0$), in which case we replace the
state label $q$ by the two-dimensional label ${\mathbf
k}=(k_1,k_2)$ (eigenvectors $\varphi_{\mathbf k}({\mathbf n})$ and
energies $E_{\mathbf k}$), as explained in Sec.~\ref{model}. The
advantage to do this will become clear below. On this basis,
Eq.~(\ref{Snm}) takes the form
\begin{equation}
S(\omega)=-\frac{1}{\pi}\mathrm{Im}\sum_{\mathbf{k,k'}}
X_{\mathbf{k,k'}} \langle \mathbf{k}|
\left\langle\hat G(\omega)\right\rangle|\mathbf{k'}\rangle\,,
\end{equation}
where we defined
\begin{equation}\label{Xkk}
X_{\mathbf{k,k'}}=\sum_\mathbf{n,m}
X_\mathbf{n,m}\varphi_\mathbf{k}(\mathbf{n})
\varphi_\mathbf{k'}^*(\mathbf{m}).
\end{equation}

We now take advantage of the fact that within the CPA the averaged
Green's function $\left\langle\hat G(\omega)\right\rangle$ is
replaced by the Green's function of the same system in the absence
of disorder, but with an $\omega$ dependent and ${\mathbf k}$
independent complex self-energy $\Sigma(\omega)$ added to the
exciton energies $E_{\mathbf k}$. Thus, within the CPA
$\left\langle\hat G(\omega)\right\rangle$ is by definition
diagonal in the $\mathbf{k}$ basis. Hence,
\begin{eqnarray}\label{Spectra}
\nonumber S(\omega) && = -\frac{1}{\pi}\mathrm{Im}\sum_{\mathbf{k}}
X_{\mathbf{k,k}} \langle \mathbf{k}|
\left\langle\hat G(\omega)\right\rangle|\mathbf{k}\rangle \\
&& = -\frac{1}{\pi}\mathrm{Im}\sum_{\mathbf{k}} X_{\mathbf{k,k}}
\frac{1}{\omega-E_\mathbf{k}-\Sigma(\omega)+{i}\eta}.
\end{eqnarray}
We proceed by using the Bloch nature of the eigenvectors
$\varphi_{\bf k}({\bf n})$ in the ring direction,
\begin{equation}
\label{Bloch} \varphi_{\mathbf k}({\mathbf n})=(N_2)^{-1/2}\exp[i2
\pi k_2 n_2/N_2] \varphi_{k_1}(n_1;k_2)
\end{equation}
with the integer $k_2$ the transverse wave number and
$\varphi_{k_1}(n_1;k_2)$ the longitudinal wave function.\cite{DIDR}
Then the strengths $X_{\mathbf{k,k}}^{A}$,
$X_{\mathbf{k,k}}^{LD}$, and $X_{\mathbf{k,k}}^{CD}$ for the three
spectra considered can be expressed in terms of the
$\varphi_{k_1}(n_1;k_2)$. This algebra was already performed in
Ref.~\onlinecite{DIDR}, yielding the oscillator strength
$O_{\mathbf k}$, the LD strength $L_{\mathbf k}$, and the
rotational strength $R_{\mathbf k}$. Combining all these
expressions, we recover for the final CPA results Eqs.~(20), (24),
and (31) of Ref.~\onlinecite{DIDR} for the absorption, the LD, and
the CD spectrum, respectively, except that
$\delta(\omega-E_{\mathbf k})$ in these expressions is replaced by
$-\frac{1}{\pi}{\text {Im}}
(\omega-E_\mathbf{k}-\Sigma(\omega)+{i}\eta)^{-1}$. In
particular we find that, like in the homogeneous case, only terms
with $k_2=0, \pm 1$ contribute to the three spectra, with the
$k_2=0$ states having a transition dipole along the $z$ axis and
the other two (degenerate) bands having a dipole perpendicular to
it. In contrast to the homogeneous case, however, to calculate the
CPA spectra we still need the eigenenergies of all states in all
$k_2$ bands. The reason is that these energies occur in the
quantity $g_0(\omega)$ (Eq.~(\ref{approximation})), which is
needed to solve for the self-energy.

We finally mention that within the CPA, the density of states is
simply given by Eq.~(\ref{Spectra}) with $X_{\mathbf{k,k}}^{\rho}$
replaced by 1.

Since the numerical diagonalization of the $N_2$ effective
one-dimensional problems that yield the longitudinal
eigenfunctions $\varphi_{k_1}(n_1;k_2)$ and the energies
$E_{\mathbf k}$ is straightforward, the only remaining problem is
to determine the self-energy $\Sigma(\omega)$. Within the CPA,
$\Sigma(\omega)$ obeys a self-consistency equation, derived by
neglecting correlations between scattering events on different
molecules. The resulting self-consistency equation
reads\cite{RICKAYZEN,ECONOMOU}
\begin{equation}\label{tmatrix}
\left\langle\frac{\epsilon_{\mathbf n}-\Sigma(\omega)}{1-
(\epsilon_{\mathbf n}-\Sigma(\omega))\langle\hat
G(\omega)\rangle_{\mathbf {nn}}} \right\rangle =0\,,
\end{equation}
where the disorder average now is a simple integration over
$\epsilon_{\mathbf n}$, weighted by the distribution ${\mathcal
P}(\omega_{\mathbf n})$. The site diagonal element of the averaged
Green's function is
\begin{equation}
\label{Gnn} \left\langle\hat G(\omega)\right\rangle_{\mathbf
{nn}}=\sum_{\mathbf k}\frac{\left| \varphi_{\mathbf k}({\mathbf
n})\right|^2}{\omega- E_{\mathbf
k}-\Sigma(\omega)+{i}\eta}
\end{equation}

The CPA has been developed for large systems, where periodic
boundary conditions can safely be applied to obtain the
homogeneous solutions.  In that case we have
$\left|\varphi_{\mathbf k}({\mathbf n})\right|^2=1/N$, with $N$ the
total number of molecules in the system. Substituting this in
Eq.~(\ref{Gnn}), we observe that the diagonal element of the
Green's function becomes site-independent, which in fact is
necessary in order for the self-energy following from
Eq.~(\ref{tmatrix}) to be site-independent. The latter fact is
implicit in the CPA by assuming from the very beginning that the
self-energy does not depend on ${\bf k}$. Conversely, if in
Eq.~(\ref{Gnn}) we would boldly substitute the solution for the
homogeneous solution with open boundary conditions applied along
the $n_1$ direction, we would obtain a site-dependent ($n_1$
dependent) self-energy, which would be inconsistent in the context
of the CPA.

Yet, as we will be interested in studying the effect of the
cylinder length on the spectra, we prefer not to use periodic
boundary conditions in the $n_1$ direction. One way out of this
impasse is to construct artificially a diagonal element of the
Green's function which is site invariant. This can be done by
approximating $\langle\hat G(\omega)\rangle_{\mathbf {nn}}$ by its
mean
\begin{eqnarray}\label{approximation}
 \left\langle\hat G(\omega)\right\rangle_{\mathbf {nn}} &\approx &
\frac{1}{N}\sum_{\mathbf n}{\left\langle\hat
G(\omega)\right\rangle_{\mathbf {nn}}}\\
\nonumber &=& \frac{1}{N}\sum_{\mathbf
k}\frac{1}{\omega- E_{\mathbf k}-\Sigma(\omega)+{\it i}\eta}\equiv
g_0(\omega),
\end{eqnarray}
which implies in particular that the local density of states is
approximated by the normalized total one. We note that in spite of
this approximation, the self-energy still contains information
about the system's finite size, as the finite-size energies
$E_{\mathbf k}$ will be used when evaluating $g_0(\omega)$.
Moreover, when calculating the spectra and density of states
(Eq.~(\ref{Spectra}) with the proper $X_{\mathbf {k,k}}$) again we
will use the exciton energies $E_{\mathbf k}$ as well as the
strength $X_{\mathbf {k,k}}$ calculated for the finite homogeneous
system. Thus, one may hope that the approximation
Eq.~(\ref{approximation}) only affects the results for the spectra
and the density of states in a rather weak way. In
Sec.~\ref{results} we will check the validity of the approximation
by comparing this new application of the CPA directly to exact
numerical simulations as well as to its traditional application,
which uses periodic boundary conditions.

We end this section by noting that solving Eq.~(\ref{tmatrix}) for
the self-energy usually requires numerical schemes. Our approach
is to rewrite this equation in the form
\begin{eqnarray}\label{iterativ}
\Sigma(\omega) &=& \left[\int {\rm d}x \, {\mathcal P}(x)\frac{x}{1-(x-
\Sigma(\omega))g_0(\omega)}\right] \\
\nonumber & & \times\left[\int {\rm d}x \, {\mathcal
P}(x)\frac{1}{1-(x-\Sigma(\omega))g_0(\omega)}
 \right]^{-1},
\end{eqnarray}
which may be solved iteratively, using as starting value
$\Sigma_0(\omega)=0$, the value of $\Sigma(\omega)$ for the
homogeneous system. The solution $\Sigma(\omega)$ to this equation
in combination with the results described below Eq.~(\ref{Bloch})
determines our CPA results for the spectra and density of states.

\section{Numerical Results for the Spectra}\label{results}

In this paper, we restrict ourselves to the application of our
formalism to the cylindrical aggregates occurring in the
chlorosomes of the bacterium {\emph{Chloroflexus
aurantiacus}}.\cite{HOLZWARTH,PROKHO,BALABAN,ROSS}  The application
to the bilayer synthetic aggregates of carbocyanine dyes is
discussed elsewhere.\cite{C8S3} In terms of our stack of rings
representation, the chlorosomal cylinders, consisting of tens of
thousands of bacteriochlorophyll {\em c} molecules, have the
following model parameters (see
Refs.~\onlinecite{PROKHO,PROKHO_NEW} and our discussion in
Ref.~\onlinecite{DIDR}): $N_2=6$, $h=0.216\,$nm,
$\alpha=189.6^\circ$, $\beta=36.7^\circ$, and $\gamma=20^\circ$.
The radius is given by $R=2.297\,$nm, while the length may extend
to hundreds of nanometers. Finally, we use an average
single-molecule transition frequency $\omega_0$ that agrees with a
wavelength of $660\,$nm,\cite{HOUSSIER,FRIGAARD,DUDKOWIAK,UMETSU}
while for the dipole squared of a single molecule we have used
$\mu^2 \approx 20~\text{Debye}^2$.  The latter value was obtained
from the integrated extinction coefficient of monomeric solutions
of BChl {\emph c}, \cite{FRIGAARD,DUDKOWIAK,UMETSU} using the
expression from Ref.~\onlinecite{HOUSSIER}.

Figure 2 presents the optical spectra for several cylinder lengths
$N_1$ calculated using these model parameters and converted to a
wavelength scale to facilitate the comparison to experiment.\cite{extranote}
All intermolecular
dipole-dipole interactions were accounted for in these spectra.
The solid lines represent the results obtained by numerical
simulation, where we averaged over $1000$ disorder realizations and
used the rigorous smoothening technique proposed by Makhov {\it et
al.}\cite{MAKHOV1,MAKHOV2} to reduce the noise in the spectra. For
the disorder strength we used $\sigma=600$ cm$^{-1}$, a value that
was chosen such that the theoretical absorption linewidth agrees
with the one observed experimentally at room temperature. Also
plotted in Fig.~2 (dashed lines) are the results obtained using
the CPA modified for open boundary conditions, as described in
Sec.~\ref{CPA}. Finally, the dotted lines represent the spectra
obtained within the traditional CPA, which assumes periodic
boundary conditions in the $n_1$ direction (at finite $N_1$
values).

       \begin{figure*}[tb]
         \centerline{
           \scalebox{0.9}{
             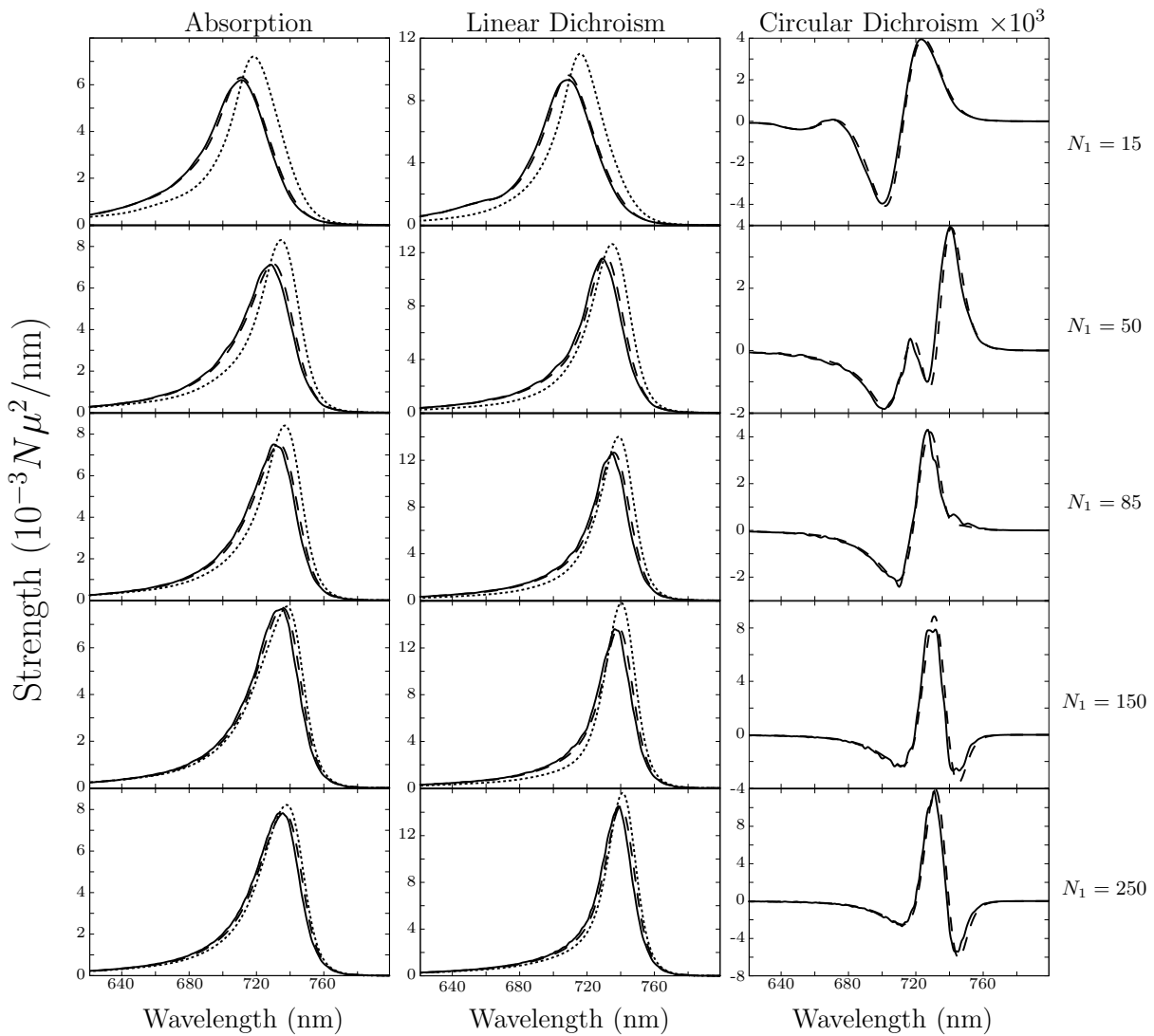
             }
           }
         \caption{Absorption, LD, and CD spectra calculated for
cylindrical aggregates with the geometry of the rod elements in
the chlorosomes of {\it{Chlorofexus aurantiacus}} (see text for details)
and a disorder strength of $\sigma=600\,$cm$^{-1}$. The solid
lines are obtained by numerical simulations (average over $1000$
disorder realizations) while the dashed lines are obtained using
the modified CPA discussed in Sec.~\ref{CPA}. The dotted lines
represent the usual implementation of the CPA, which starts from
periodic boundary conditions. From top to bottom, the cylinder
length is varied as follows: $N_1=15$, $50$, $85$, $150$, and
$250$.}
       \end{figure*}

As is clear from Fig.~2, the modified CPA is in excellent
agreement with the numerical simulations for all spectra and at
all sizes. Both shape and position of the spectral features agree
almost perfectly. We conclude that the finite-size effects in the
spectra are captured in an excellent way by the approximation made
in Eq.~(\ref{approximation}). These results justify the use of the
CPA to model the spectra of wider cylinders,\cite{C8S3} where the
size of the one-exciton space becomes so large that the
brute-force numerical simulation of the spectra becomes
computationally too expensive.

To demonstrate that the proposed modification of the CPA is in
fact essential to cover the finite-size effects, we have also
applied the CPA in the traditional way, imposing periodic boundary
conditions in the $n_1$ direction. In an attempt to still account
for finite-size effects, we have however kept the length $N_1$ of
the cylinder finite. We thus used Bloch waves for the longitudinal
wave functions $\varphi_{k_1}(n_1;k_2)$.\cite{DIDR} The results
for the absorption and LD spectra are shown as dotted curves in
Fig.~2. For the CD spectrum, the use of periodic boundary
conditions may be shown to be consistent only in the limit $N_1\to
\infty$,\cite{DIDR} which is why dotted curves are absent in the
CD panels. It is clear from Fig.~2 that the CPA with periodic
boundary conditions yields spectra that differ notably from the
exact ones, both in the position and shape of spectral features.
The agreement is especially bad for short cylinders, where
finite-size effects are most prominent. At all sizes considered
(even for $N_1=250$) the modified CPA constitutes a better
approximation than the one with periodic boundary conditions. We
note that the spectra obtained by using periodic boundary
conditions are always red-shifted relative to the exact and the
modified-CPA spectra. This results from the fact that for periodic
boundary conditions every molecule interacts with other molecules
that are at most half a cylinder length away, while using the
correct open boundary conditions, the molecules near one edge of
the cylinder ($n_1$ small) have much weaker interactions with the
molecules at the other edge ($n_1 \approx N_1$). Thus, using
periodic boundary conditions, one overestimates the effect of the
interactions, which shifts the spectrum too far away from the
monomer transition.

As final issue, we discuss the comparison between the exact
spectra and the experimental ones. In our previous work\cite{DIDR}
we made this comparison neglecting the effects of disorder in the
model. One may consider those previous results as one extreme
case, where all spectral broadening is assumed to arise from
homogeneous broadening, while the current spectra constitute the
other extreme situation, where the widths are assumed to purely
result from inhomogeneity. Like in the homogeneous case reported
in Ref.~\onlinecite{DIDR}, we see that in particular the CD
spectra exhibit a strong dependence on the length $N_1$ of the
cylinder. This length dependence, specifically the change of the CD
line shape around $N_1=100$, was an important point in our previous
work, as it suggests that the strong variation in the reported CD
spectra of
chlorosomes\cite{GRIEBENOW,LEHMANN,WANG,FRESE,STEENSGAARD} results
from the fact that different samples contain chlorosomes of
different length. Apparently this conclusion survives the
incorporation of disorder and the concomitant localization of the
exciton states.

We stress that at the disorder value considered the typical
exciton localization size in the region of the absorption band is
several tens of molecules (Sec.~\ref{results2}). Thus, one would
expect the spectra to be size-saturated at cylinder lengths of at
most several tens of rings. For the absorption and LD spectra,
this indeed is the case, except that small shifts of the entire
line shape still occur for longer cylinders as a result of the
long-range dipole-dipole interactions. The slower size saturation
for the CD spectrum results from two aspects. First, the presence
of the intermolecular distances in the expression for the
rotational strengths (cf.~Eq.~(\ref{XCD})) contribute to a
prolonged size dependence. Second, being a difference spectrum,
the CD spectrum is much more sensitive to the already mentioned
small shifts in the exciton energies that result from the
long-range interactions. The fact that for clorosomes the length
dependence of the CD spectrum survives the inclusion of disorder
was also suggested by Prokhorenko et al.\cite{PROKHO_NEW} They
based this conclusion on a study of the so-called CDM-matrix over
a limited length interval, rather than a direct study of the
spectrum.

While the spectra for the model with disorder follow the same
general trends as those without disorder, the more detailed
comparison to experiment is better for the case with disorder.
First, in the presence of disorder the high-energy dip in the CD
spectra for $N_1 > 100$ is seen to have a smaller amplitude than
the low-energy dip, while this ratio is opposite for the
homogeneous case.\cite{DIDR} In experiment, the type of CD spectra
with two negative dips indeed always have a smaller amplitude for
the high-energy dip. The effect of disorder is to smear out the
high-energy dip, giving it a smaller amplitude. Second, the
presence of disorder gives the absorption and LD spectra a more
pronounced high-energy tail than is obtained for the homogeneous
model. Indeed, these tails, quite typical for disordered J
aggregates, are observed in experiment.

To finish this section, we present in Fig.~3 the density of states
for the disordered model (solid line, obtained from simulations)
and the homogeneous one (dotted line), both for $N_1=250$. As
above, the disorder was taken to be $\sigma=600$ cm$^{-1}$; the
sticks in the density of states for the homogeneous case were
convoluted with Lorentzian curves of FWHM$=20\,$cm$^{-1}$. The
disorder clearly smears the discrete peaks still visible for the
homogeneous case and leads to one broad feature that peaks
somewhere in the middle of the exciton band. This is in marked
contrast to the frequently studied one-dimensional aggregates,
where the density of states, even in the presence of disorder,
peaks at the band edges. So, even while the cylinder constitutes a
strongly anisotropic system, its density of states deviates from a
simple one-dimensional picture.

       \begin{figure}[tb]
         \centerline{
           \scalebox{1.0}{
             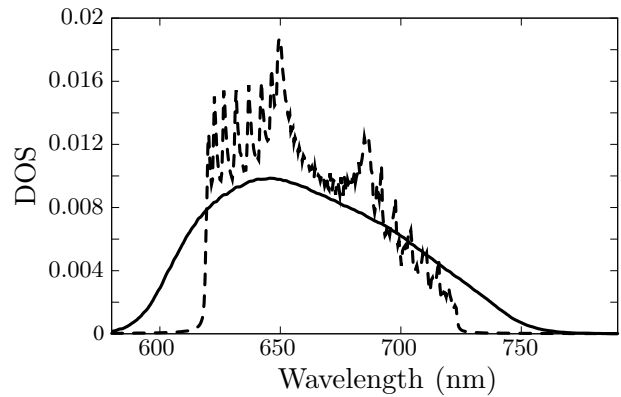
             }
           }
         \caption{Density of states for homogeneous (dashed line)
and disordered (solid line) cylindrical aggregates of the
chlorosome structure with a length of $N_1=250$ rings. In the
former case, the excitons were homogeneously broadened by a
Lorentzian of FWHM 20 cm$^{-1}$, while in the latter case a
disorder strength of $\sigma=600\,$cm$^{-1}$ was used. The disorder
results were obtained from numerical simulations, averaging over
$1000$ realizations.}
       \end{figure}

\section{Numerical Results for the Localization Characteristics}\label{results2}

We now turn in more detail to the localization of the exciton wave
functions. As a first step, we consider the participation ratio
${\mathcal L}^{-1}(\omega)$. In Fig.~4, we have plotted
$9{\mathcal L}^{-1}(\omega)/4$ for chlorosomes within the
homogeneous model (Fig.~4(a), $N_1=250$) and the disordered model
(Fig.~4(b), $N_1=150$, $200$, $250$, and $300$). In the homogeneous model
we replaced the delta functions in Eq.~(\ref{PARRATIO}) and the
density of states by Lorentzians with a FWHM of $20\,$cm$^{-1}$,
while in the disordered case (disorder strength $\sigma=600$
cm$^{-1}$, as in the previous section), we used the smoothening
technique\cite{MAKHOV1,MAKHOV2} to reduce the noise in the
simulations. The normalization factor $9/4$ was introduced to
guarantee that for the homogeneous case we recover the total
number of molecules $N$ in the cylinder, see discussion below
Eq.~(\ref{PARRATIO}). This is clear from Fig.~4(a), where, indeed,
inside the exciton band $9{\mathcal L}^{-1}(\omega)/4$ obtains an
almost constant value of 1500 molecules, the total number of
molecules in a cylinder of 250 rings.

       \begin{figure}[tb]
         \centerline{
           \scalebox{1.0}{
             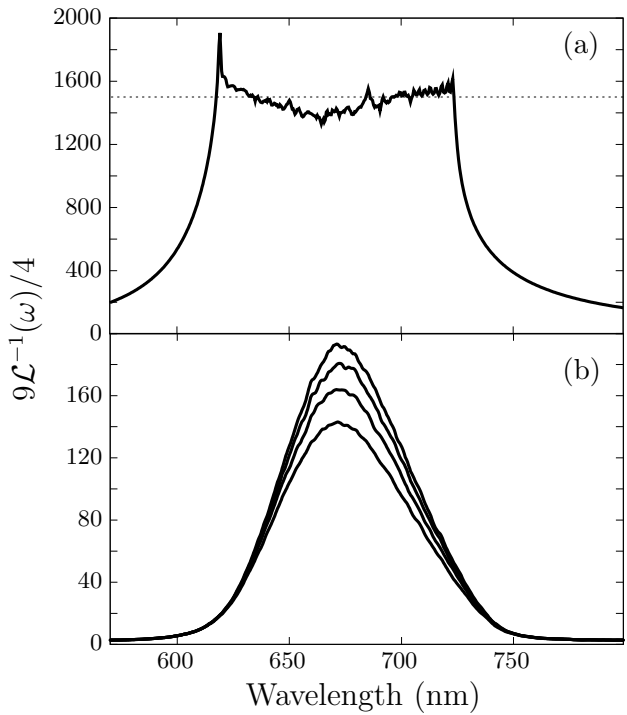
             }
           }
         \caption{(a) Energy dependent participation ratio for
homogeneous cylindrical aggregates of the chlorosome structure
with a length of $N_1=250$ rings. The dotted line indicates the total
number of molecules (1500) in the cylinder. 
(b) As in panel (a), but now in
the presence of diagonal disorder of strength
$\sigma=600\,$cm$^{-1}$ and considering four cylinder lengths.
From top to bottom the curves correspond to $N_1=300$, $250$,
$200$, and $150$, respectively.}
       \end{figure}

In Fig.~4(b) we see that the disorder strength of $600\,$cm$^{-1}$
leads to clear localization of the exciton states at all energies,
even in the centre of the band, where the robustness against
localization always is strongest. The fact that the band edge
states are rather strongly localized is not surprising: the ratio
of the disorder strength $\sigma$ and the total exciton bandwidth
is approximately $0.26$. In the spectral region where the
absorption band occurs ($720-750\,$nm), the participation ratio
yields a number of several tens of molecules over which the
exciton states are delocalized. We emphasize that this number
varies substantially (by a factor of $8$) over the width of the
absorption band. On the other hand, it is seen that in this energy
region, the participation ratio hardly depends on the cylinder
length anymore, which is in accordance with the fact that the
exciton states are localized on a region of the cylinder that is
much smaller than its total size. At the peak of the absorption
spectrum (about $740\,$nm), the calculated participation ratio
implies that the excitons are shared coherently by about $15$
molecules (the autocorrelation function yields $11$ molecules, see
below). This value may be compared with the number of $7.4-7.6$
obtained from measuring the bleaching ratio in absorption
difference experiments.\cite{SAVIKHIN,YAKOVLEV} Given the large
variation in the localization size over the absorption band, the
arbitrariness and uncertainty present in any definition of a
localization size, and the nonlinear nature of the absorption
difference experiments, the agreement between our simulations and
these experimental data is good. Our localization size is
considerably smaller than the $40-50$ molecules obtained in the
simulations of Prokhorenko et al.\cite{PROKHO_NEW} Still, for the
same reasons as stressed above already, considerable room exists
for deviations in the values reported from different model studies.

To obtain insight into the possible anisotropy of the localization
properties, we also studied the autocorrelation function
${\mathcal C}({\mathbf n};\omega)$ defined in Eq.~(\ref{COH}). We
note that this anisotropy may also be studied by plotting
individual wave functions, as we did in Fig.~10 of
Ref.~\onlinecite{DIDR04}, but such plots have the drawback that
they represent arbitrarily picked and {\em hopefully} typical
states, while the correlation function gives statistical
information. In Fig.~5 we present three-dimensional plots ((a)-(c))
as well as contour plots ((d)-(f)) of ${\mathcal C}({\mathbf
n};\omega)$ for frequencies corresponding to wavelengths of,
respectively, $700\,$nm, $740\,$nm, and $780\,$nm for a cylinder of length
$N_1=250$ and a disorder strength of $\sigma=600$ cm$^{-1}$. To
make these plots, the cylinder was cut along a line parallel to
the $z$ axis (the $n_1$ axis in the plots) and unwrapped.

       \begin{figure*}[tb]
         \centerline{
           \scalebox{0.9}{
             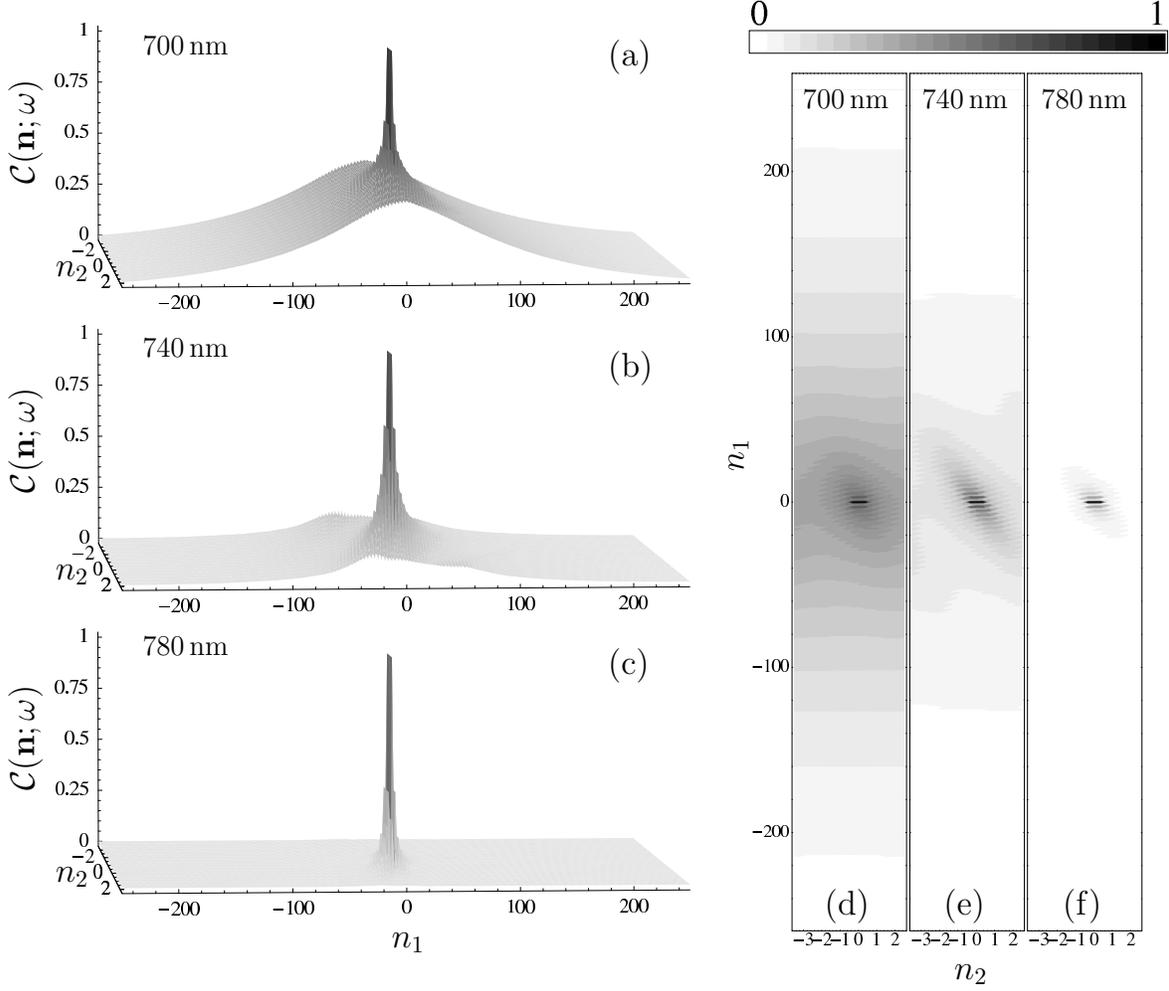
             }
           }
         \caption{Three-dimensional plots ((a)-(c)) and contour
plots ((d)-(f)) of the autocorrelation function ${\mathcal
C}({\mathbf n};\omega)$ [Eq.~(\ref{COH})] for cylindrical
aggregates of the chlorosome structure with a length of $N_1=250$
rings and a disorder strength of $\sigma=600\,$cm$^{-1}$ (averaged
over $150$ realizations) at three different energies,
corresponding to $700\,$nm ((a) and (d)), $740\,$nm ((b) and (e)),
and $780\,$nm ((c) and (f)). The cylinder surface is represented
by cutting it along a line parallel to the $z$ axis and unwrapping
it. This cutting line is the $n_1$ axis in the plots; points with
constant value of $n_1$ lie on the same ring of the cylinder. Due
to the helical structure of chlorosomes, lines of constant $n_2$
make a finite angle with the $n_1$ direction (compare the dashed
line in Fig.~1). The contour plots distinguish 25 equally large
intervals for the value of ${\mathcal C}({\mathbf n};\omega)$,
represented on a grayscale (see legend above the contour plots).}
       \end{figure*}

Several observations can immediately be made from Fig.~5. First,
with increasing energy the wave functions clearly get more
extended, which is in agreement with Fig.~4(b). In fact, if we
calculate $N_{del}^C(\omega)$ (defined at the end of
Section~\ref{model}) from ${\mathcal C}({\mathbf n};\omega)$, we
find for $N_1=250$ the values $N_{del}^C(700\,\text{nm})=147$,
$N_{del}^C(740\,\text{nm})=11$, and $N_{del}^C(780\,\text{nm})=3$.
These values are in reasonably good agreement with those obtained
from the participation ratio: $9\mathcal{L}^{-1}(700\,\text{nm})/4
= 119$, $9\mathcal{L}^{-1}(740\,\text{nm})/4 = 15$,
and $9\mathcal{L}^{-1}(780\,\text{nm})/4 = 2.9$. We also
note that the values for $N_{del}^C(\omega)$ (and the plots for
${\mathcal C}({\mathbf n};\omega)$) at 740 and 780 nm do not depend
on the cylinder length anymore at $N_1=250$. Second, as is most
clearly visible in the contour plots, at the short length scale
the wave functions exhibit a clear anisotropy along a direction
that is neither given by $n_1$=constant, nor by $n_2$=constant,
i.e., neither in the ring direction, nor in the direction of the
helices drawn as dashed line in Fig.~1. In fact, the slanting
direction of the contours relative to the vertical axis observed
in Fig.~5 (which varies slightly with varying energy), closely
resembles the slanting of the equal-phase lines of the wave
functions in the $k_2= \pm 1$ bands of the homogeneous cylinder
discovered in Ref.~\onlinecite{DIDR04}. In the remainder of this
section, we will explain this behavior, starting with a
perturbative picture in which weak disorder mixes exciton states
within and between the $k_2$ bands that exist for homogeneous
cylinders.

In Ref.~\onlinecite{DIDR04}, we have shown that the optically
dominant exciton states in the $k_2=0$ and $k_2=\pm 1$ bands of the
homogeneous cylinder are well approximated by analytical
expressions of the form Eq.~(\ref{Bloch}) with the ansatz
\begin{equation}\label{ansatz}
\varphi_{k_1}(n_1;k_2)= \sqrt{\frac{2}{N_1+1}}\sin \left(\frac{\pi
k_1 n_1}{N_1+1}\right) {\mathrm e}^{i s(\theta_{k_2}
+|k_2|\gamma)n_1}.
\end{equation}
Here, $k_1$ is a positive integer, small compared to $N_1$ for the
states of interest, $s=k_2/|k_2|$, $\gamma$ is the helical angle
of the cylinder (see Fig.~1), and $\theta_{k_2}$ is a phase angle
that is used to optimize the quality of the ansatz for the complex
effective one-dimensional Hamiltonian of the band with wave number
$k_2$.\cite{DIDR04} The ansatz Eq.~(\ref{ansatz}) is exact if
transfer interactions only occur between molecules on
nearest-neighbor rings; in that case, the total angle
$\theta_{k_2} +|k_2|\gamma$ is the phase of the effective
nearest-ring interaction. Optimizing $\theta_{k_2}$ approximately
accounts for mixing of the wave functions Eq.~(\ref{ansatz}) by
non-nearest-neighbor interactions. The angle $\theta_{k_2}$ was
found to be responsible for the slanting of equal-phase lines of
the wave functions in the $k_2=\pm 1$ band of the homogeneous
cylinder (chirality of the wave functions) relative to the
vertical axis, which seems essential to understand the behavior of
${\mathcal C}({\mathbf n};\omega)$ observed in Fig.~5(d)-(f). We
therefore reconsider the ansatz solutions Eq.~(\ref{ansatz}), with
a special focus on $\theta_{k_2}$ and extending the treatment to
the bottom regions of all $k_2$ bands.

{\em The $k_2=0$ band.} The Hamiltonian for the $k_2=0$ band is
real, implying that also the wave functions can be chosen real.
Hence, we have $\theta_{0}=0$. It was shown in
Ref.~\onlinecite{DIDR04} that the bottom state of this band (the
$k_1=1$ state, which carries $81\%$ of the oscillator strength in
this band) is well described by the ansatz. For this band, the wave
function has equal amplitude on all molecules of a certain ring,
while the quantum number $k_1$ gives the number of maxima along
the $z$ axis of the cylinder. This is illustrated in Fig.~6(a),
where we plotted the amplitude squared of the exact wave function
(obtained by numerical diagonalization) of the lowest state in the
$k_2=0$ band of the homogeneous cylinder.

       \begin{figure}[tb]
         \centerline{
           \scalebox{1.0}{
             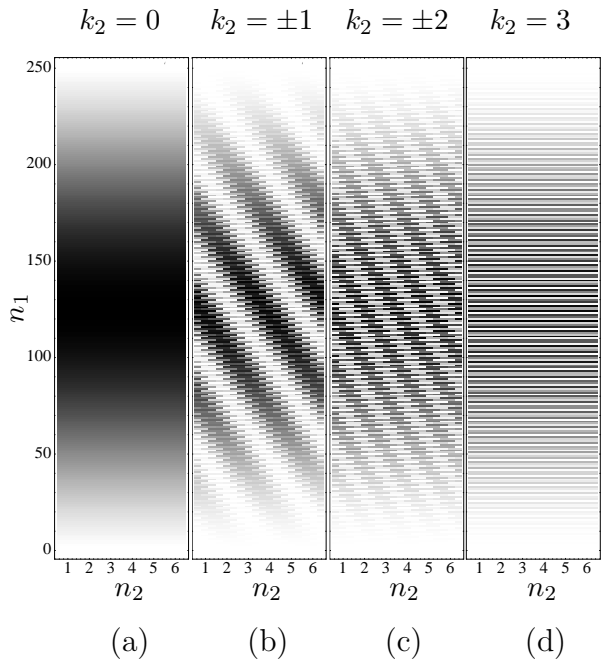
             }
           }
         \caption{Gray-scale density plots of the squared
amplitudes (darker represents a higher amplitude) of the lowest
states in the six exciton bands of homogeneous cylindrical
aggregates of the chlorosome structure with a length of $N_1=250$
rings: $k_2=0$ (a), $k_2=\pm 1$ (b), $k_2=\pm 2$ (c), and $k_2=3$
(d). Plots are given on the unwrapped cylinder surface. The wave
functions were obtained from exact diagonalization of the
corresponding homogeneous Hamiltonians and are well described by
Eq.~(\ref{Bloch}), with the ansatz Eq.~(\ref{ansatz}) (see text
for details). Relative to the average molecular transition energy
$\omega_0$, the lowest states have the energies
$-1288.73\,$cm$^{-1}$ (a), $-1291.92\,$cm$^{-1}$ (b),
$-1244.97\,$cm$^{-1}$ (c), and $-1166.09\,$cm$^{-1}$ (d).}
       \end{figure}

{\em The $k_2=\pm 1$ bands.} The effective Hamiltonian for these
two degenerate bands is essentially complex and the phase factor in
Eq.~(\ref{ansatz}) is needed. In Ref.~\onlinecite{DIDR04} we have
shown that the superradiant states (three in each band) all occur
near $k_1=k^*={\mathrm{nint}[|\theta_1|(N_1+1)/\pi]}$ (nint
denoting the nearest integer function), which is a finite energy
above the bottom of the $k_1=\pm 1$ bands. Excellent agreement was
found between the ansatz wave functions and those for the
superradiant states of the homogeneous cylinder obtained by
numerical diagonalization for the optimized value of $\theta_{\pm
1}=4.3^\circ$.\cite{DIDR04} Presently, we are not only interested
in the superradiant states, but even more so in the states near
the bottom of the $k_2=\pm 1$ bands ($k_1$ in the order of unity),
because these states are closest to the bottom of the $k_2=0$ band
and should be expected to mix with those states in the presence of
(weak) disorder. We may apply the same approach as followed in
Ref.~\onlinecite{DIDR04} for the superradiant states to the states
near the band bottom. Thus, we fix $\theta_{\pm 1}$ such that the
mixing between the bottom states in the $k_2=\pm 1$ bands
resulting from the long-range interactions does not diverge. This
eventually yields $\theta'_{\pm 1}=3.8^\circ$ (for $N_1=250$),
which only slightly differs from the phase angle in the
superradiant region.

If we plot the amplitude squared of the exciton wave function on
the unwrapped cylinder, a state of the form Eq.~(\ref{Bloch}) with
Eq.~(\ref{ansatz}) gives lines of equal intensity that are dictated
by lines of equal phase $s(\theta_{k_2} +|k_2|\gamma)n_1+2\pi k_2
n_2/N_2$. The angle $\chi$ of these lines with the $z$ axis is
easily derived to obey
\begin{equation}
\tan {\chi}=-\frac{\theta_{k_2} R}{|k_2|h}.
\end{equation}
This implies that the wave function rotates around the cylinder
over a number rings given by
\begin{equation}\label{complete_rotation}
n_1^*=2\pi |k_2/\theta_{k_2}|.
\end{equation}
From this we find that in $k_2=\pm 1$ bands $n_1^* \approx 84$ for
the wave functions of the superradiant states and $n_1^* \approx
95$ for the bottom states. In Fig.~6(b) we show the amplitude
squared for the bottom state of the $k_2=1$ band obtained by
numerical diagonalization. Of course, on every ring we find a
modulation with two maxima, as is appropriate for this band. More
importantly, we find good agreement between the value of $n_1^*$
that may be obtained from this plot and the above estimate
obtained from the ansatz wave function.

{\em The $k_2=\pm 2$ bands.} These bands, which contain no
oscillator strength in the homogeneous limit, were not considered
in Ref.~\onlinecite{DIDR04}. They may be treated in a way similar
to the $k_2=\pm 1$ bands. If we focus on optimizing the ansatz at
the bottom of the bands, we find $\theta_{\pm 2}=6.6^\circ$ for
$N_1=250$ and that for this value the ansatz Eq.~(\ref{ansatz})
for $k_1=1$ indeed gives a good description of the numerically
obtained lowest state.  From Eq.~(\ref{complete_rotation}) and
$\theta_{\pm 2}=6.6^\circ$ a value of $n_1^*\approx 109$ can be
estimated, which is seen to be in good agreement with the exact
wave function plotted in Fig.~6(c). Of course, the states in the
$k_2=\pm 2$ bands exhibit a modulation of squared amplitudes
inside each ring with four maxima.

{\em The $k_2=3$ band.} This band, too, has no oscillator
strength. Like the $k_2=0$ band, the $k_2=3$ band is nondegenerate
and is governed by a real Hamiltonian. Thus, $\theta_3+3\gamma=0$
and, like for the $k_2=0$ case, no optimization condition for
$\theta_3$ has to be solved. Figure 6(d) gives the amplitude
squared of the numerically obtained lowest state in this band,
which exhibits no modulation inside rings (the Bloch factor is
$(-1)^{n_2}$ which gives unity upon taking the square). One does
observe a vertical modulation with a high periodicity, which is
due to the fact that for this band the lowest state is not the
$k_1=1$ state, but rather a state that corresponds to a high value
of $k_1$. This is due to the fact that for this band the effective
Hamiltonian has dominant positive in stead of negative
interactions.

       \begin{figure}[tb]
         \centerline{
           \scalebox{1.0}{
             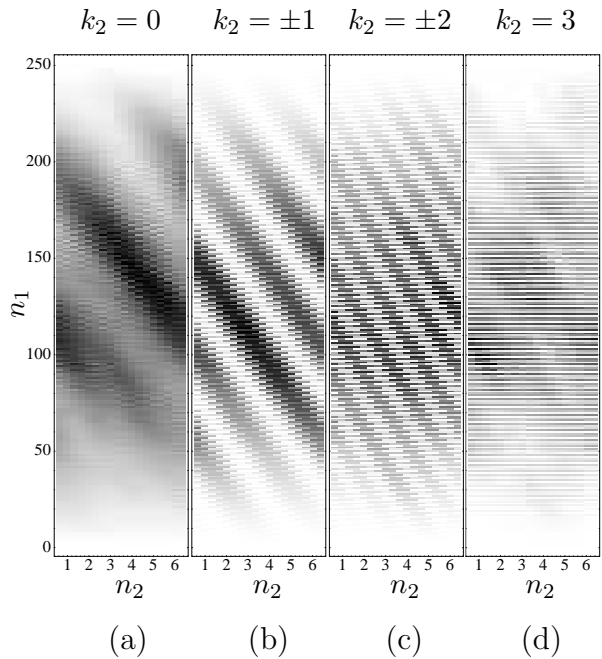
             }
           }
         \caption{As in Fig.~6, but now including diagonal
disorder of a strength $\sigma=20\,$cm$^{-1}$. The weakness of the
disorder allows for an identification to the ${\bf k}$ states of
the homogeneous aggregate in Fig.~6. The energies of the plotted
states are $-1288.96\,$cm$^{-1}$ (a), $-1292.67\,$cm$^{-1}$ (b),
$-1245.97\,$cm$^{-1}$ (c), and $-1167.55\,$cm$^{-1}$ (d).}
       \end{figure}

The important conclusion of the above analysis is that the bottom
states in the $k_2=0$ and $k_2=3$ bands exhibit no chiral behavior
(no equal-intensity lines slanted relative to the $z$ axis), while
the bottom states of the $k_2=\pm 1$ and the $k_2=\pm 2$ bands do
show a chirality which, moreover, is very similar in magnitude as
estimated by the number $n_1^*$. If we now allow for weak
disorder, mixing of the ${\bf k}$ states within and between bands
will occur. Focusing on the bottom region of the density of states,
we thus find that the wave functions in the homogeneous $k_2=0$ and
$k_2=3$ bands will (perturbatively) acquire a slanted contribution
as well. In view of the fact that the slant angle varies little
inside as well as between the $k_2=\pm 1$ and $k_2=\pm 2$ bands,
we thus expect a similar slanting for all low-energy states at
weak disorder. This is clearly confirmed by Fig.~7, where we
presented the bottom states of the $k_2=0$, 1, 2, and 3 bands
weakly perturbed by the presence of a very small disorder strength
of $\sigma=20$ cm$^{-1}$.  If we further increase the disorder
strength, the mixing will become nonperturbative and will also
involve higher-lying states, described by different $\chi$ values
and a different number of nodes in the vertical direction. These
are the ingredients for the localization visible in Fig.~5;
however, the slanting of the autocorrelation function still
clearly reflects the fact that locally the wave functions have a
chirality that derives from the homogeneous states. The reason is
that this chirality is mainly driven by the dominant transfer
interactions, which occur over a few rings.

\section{Conclusions}\label{conclusions}

In this paper we have theoretically investigated the effects of
disorder on the linear optical properties and the localization
behavior of the exciton states of cylindrical molecular aggregates.
As specific example, we have used the structure for the cylindrical
aggregates found in the light harvesting systems (chlorosomes) of
green bacteria. We have calculated the absorption, LD, and CD
spectra for various cylinder lengths in the presence of Gaussian
diagonal disorder, using both numerical simulations and the CPA.
To this end, we modified the usual implementation of the CPA to
account for finite-size effects (open boundary conditions). By
comparison to the simulation results, we have shown that this new
implementation yields an excellent approximation, in fact
significantly better than the usual CPA, which starts from
periodic boundary conditions. We have shown that for the
chlorosomes the inclusion of disorder improves the comparison to
experiment in several respects. Most importantly, however, the
inclusion of disorder does not affect the main conclusion of
Ref.~\onlinecite{DIDR} that the CD spectrum has a strong size
dependence up to cylinders of hundreds of rings long. We
demonstrated and discussed that this effect occurs in spite of the
fact that the exciton localization size is far smaller than the
cylinder size.

The localization behavior of the excitons was studied by examining
two quantities: the participation ratio and an autocorrelation
function of the exciton wave functions. While the former only
yields an approximate measure for the number of molecules
participating in the excitation, the latter gives additional
information about the direction of localization on the cylinder.
In the case of chlorosomes, we found that the excitons that
dominate the optical properties (the ones near the band bottom)
have a strongly anisotropic localization behavior, being extended
mainly along helices, whose direction is dictated by the interplay
of the various intermolecular excitation transfer interactions in
the cylinder. As we have demonstrated, this chiral behavior finds
its roots in the chirality of the fully extended exciton states
for the homogeneous cylinder. Both the participation ratio and the
autocorrelation function show that, starting from the bottom of
the exciton band, the exciton states become more extended with
increasing energy (both quantities give comparable values for the
energy dependent localization size). Also the chirality of the
wave functions is energy dependent, but this effect is small. It
is of interest to speculate whether the chiral nature of the wave
functions could be detected in terms of a rotating polarization of
the light emitted by an exciton while propagating. This would
require single-aggregate experiments with high spatial resolution
and a time-resolved detection.\cite{KOEHLER03,YU}

\section* {Acknowledgment}

This work is part of the research program of the Stichting voor
Fundamenteel Onderzoek der Materie (FOM), which is financially
supported by the Nederlandse Organisatie voor Wetenschappelijk
Onderzoek (NWO).


\begin{thebibliography}{60}
\bibitem[{foo()}]{footnote}
\bibinfo{note}{The states $\mathbf{k}=(0,0)$ and $(N_1/2,N_2/2)$ have a
  participation ratio $1/N$, while a participation tario of $3/2N$ is obtained
  for the states $(0,k_2)$, $(k_1, 0)$, $(N_1/2,k_2)$, and $(k_1,N_2/2)$, with
  $k_i \not\in \{0,N_i/2\}$; Of course the states $N_i/2$ exist only when $N_i$
  is even.}
\end{thebibliography}


\end{document}